\documentclass[prb,aps,twocolumn,amsmath]{revtex4}
\usepackage{graphicx}
\usepackage{epsfig}
\begin{document}
\title{Non-magnetic semiconductor spin transistor}
\author{K. C. Hall, Wayne H. Lau, K. G\"{u}ndo\u{g}du, Michael E. Flatt\'{e}, and Thomas F. Boggess}
\affiliation{Department of Physics and Astronomy and Optical Science and Technology Center, The University of Iowa, Iowa City,\\
Iowa 52242}
\begin{abstract}
We propose a spin transistor using only non-magnetic materials
that exploits the characteristics of bulk inversion asymmetry
(BIA) in (110) symmetric quantum wells. We show that extremely
large spin splittings due to BIA are possible in (110)
 InAs/GaSb/AlSb
heterostructures, which together with the enhanced spin decay
times in (110) quantum wells demonstrates the potential for
exploitation of BIA effects in semiconductor spintronics devices.
 Spin injection and detection is achieved using spin-dependent
resonant interband tunneling and spin transistor action is
realized through control of the electron spin lifetime in an InAs
lateral transport channel using an applied electric field (Rashba
effect). This device may also be used as a spin valve, or a
magnetic field sensor. The electronic structure and spin
relaxation times for the spin transistor proposed here are
calculated using a nonperturbative 14-band ${\bf k\cdot p}$
nanostructure model.
\end{abstract}

\maketitle

A number of semiconductor spintronic devices have been proposed in
recent years that rely on the energy splitting between electron
spin states arising from structural inversion
asymmetry,\cite{Datta:1990,Voskoboynikov:2000,deAndradaeSilva:1999,Koga:PRL2002,Ting:2002}
also known as the Rashba effect.\cite{Rashba:1984} Among these
device concepts, those involving only \emph{non-magnetic}
materials are especially attractive for application to high-speed,
spin-sensitive electronics, since they avoid the complex materials
issues and unwanted stray magnetic fields associated with the
incorporation
 of magnetic contacts, and because their operation relies on applied electric fields only,
 which
 may be modulated at considerably higher rates than magnetic fields.
 The 6.1-\AA \hspace{.02in}lattice constant family of heterostructures
(AlSb/GaSb/InAs) offers substantial advantages for such
applications because of the high electron mobility of InAs and the
extremely large spin splittings characteristic of these
heterostructures.\cite{Ting:2002,Cartoixa:2001,Grundler:2000}
However, in devices relying entirely on the Rashba effect, a
tradeoff exists between the spin splitting and the spin relaxation
time ($T_1$) due to the characteristics of the associated crystal
magnetic field. In a III-V semiconductor heterostructure, the
magnitude and direction of the wavevector-dependent crystal
magnetic field determines the size of the spin splitting and the
rate of spin relaxation through precessional
decay.\cite{Lau:2001,ALS:book,DP:1972} The Rashba effective
magnetic field lies in the plane of the heterostructure and
perpendicular to the electron
wavevector.\cite{Ting:2002,Cartoixa:2001,Rashba:1984} In this
case, regardless of the choice of the non-equilibrium spin
orientation to be used within a specific spintronics device,
promising designs incorporating a large Rashba spin splitting will
suffer from rapid precessional spin
relaxation,\cite{noteDTingsFilterT1} placing serious limitations
on feasible device architectures.

\begin{figure}[t]\vspace{0pt}
    \includegraphics[width=8.0cm]{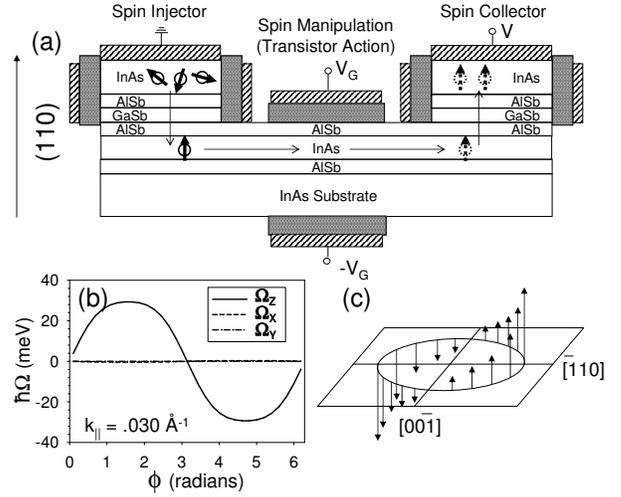}
    \caption{(a) Schematic diagram of a non-magnetic semiconductor spin transistor.  The contacts
    involving extra (shaded) layers represent metal-on-insulator gates.  The spin relaxation time in the 2DEG lateral transport
    channel is controlled using an applied electric field.
     (b) Calculated crystal magnetic field
    for HH1 in the GaSb/AlSb quantum well within the
proposed spin injector/detector for a fixed in-plane electron
wavevector (k$_{||}$) of 0.03 \AA$^{-1}$\hspace{.02in} versus
in-plane angle ($\phi$). $\phi=0$ corresponds to the $[00\bar{1}]$
direction. (c) Schematic diagram of the calculated field
    in (b).  Due to the field characteristics in (b) and (c), the side contacts on the RIT structures must be oriented with the $[\bar{1}10]$ axis, however
    the direction of transport in the InAs 2DEG is unrestricted.  Lateral transport along $[\bar{1}10]$ is shown only for simplicity of illustration.}
    \label{fig:110SpinValveAndOmega}
\end{figure}
Here we propose a spin transistor using non-magnetic materials
that exploits the unique characteristics of bulk inversion
asymmetry (BIA) in (110)-oriented semiconductor heterostructures.
Since the BIA crystal magnetic field in (110) symmetric quantum
wells is oriented approximately in the growth direction for all
electron wavevectors ({\bf k}),\cite{DK:1986} in devices based on
such structures there is a natural choice of quantization axis for
spin along which precessional spin relaxation is
suppressed.\cite{DK:1986,Ohno:1999,OurCondmat} In the present
work, we demonstrate that large BIA spin splittings are possible
in (110) InAs/GaSb/AlSb heterostructures, exceeding reported
Rasbha spin splittings in this
system\cite{Cartoixa:2001,Heida:1998,Matsuyama:2000} and in
InGaAs/InAlAs
heterostructures.\cite{Grundler:2000,Sato:2001,Nitta:1997,Gui:2000}
\begin{figure}[t]\vspace{0pt}
    \includegraphics[width=8.0cm]{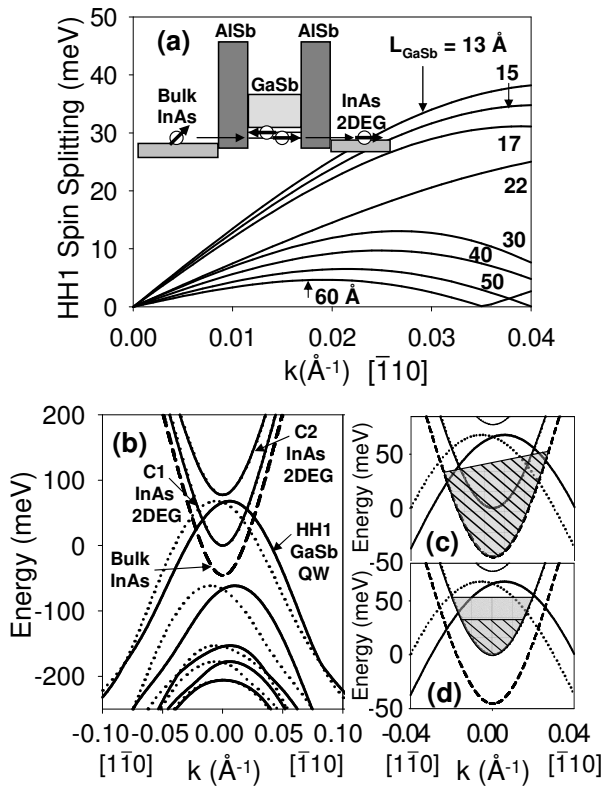}
    \caption{Description of a spin filter and detector based on RIT in a InAs/AlSb/GaSb/AlSb/InAs heterostructure.
    (a)  Calculated spin splitting in a GaSb/AlSb quantum well versus quantum well thickness (L$_{AlSb}$ = 60 \AA \hspace{.02in});
     (b)  Overlay of the calculated band structure of the 13 \AA
\hspace{.02in}GaSb/AlSb quantum well, the bulk
    InAs emitter, and the 215 \AA \hspace{.02in}InAs/AlSb quantum well forming the lateral transport channel.  For the GaSb quantum well, spin up
(down) states with respect
    to the (110) growth direction (+z) are indicated by the dotted (solid) curves.  For bulk InAs and the weakly confined 2DEG, the
    spin splittings are not resolvable on this energy scale.  The zero of energy has been chosen at the
edge of the first conduction subband (C1) of the InAs 2DEG;
(c),(d) Close-up view of the states involved in resonant
tunneling, under conditions of (c) spin
    injection and (d) spin detection.  +z (-z) polarized
electrons are indicated by the shaded (diagonal striped) region.}
    \label{fig:110SpinFilter}
\end{figure} The device, which is depicted schematically in
Fig.~\ref{fig:110SpinValveAndOmega}(a), utilizes spin-dependent
resonant interband tunneling (RIT)\cite{Ting:2002,Soderstrom:1989}
in a (110) InAs/AlSb/GaSb/AlSb/InAs heterostructure for both the
generation and detection of spin-polarized electrons. Spin
transistor action is realized through the application of an
external electric field to a lateral transport channel formed by a
high mobility, symmetric InAs two-dimensional electron gas (2DEG).
The applied field reduces the spin relaxation time in the InAs
quantum well through the electric field-induced Rashba
effect\cite{Rashba:1984,Grundler:2000,Sato:2001,Nitta:1997,Heida:1998,Matsuyama:2000,Lau:JAP2002}
yielding unpolarized carriers following lateral transport.  The
electronic structure and spin relaxation times for the spin
transistor proposed here are calculated using a nonperturbative
14-band ${\bf k\cdot p}$ nanostructure model,\cite{Lau:2001} in
which BIA is included naturally to all orders of the electron
wavevector.

The central feature exploited in the device proposed here is the
structure of the BIA crystal magnetic field in (110) symmetric
quantum wells. As shown in
Fig.~\ref{fig:110SpinValveAndOmega}(b)-(c), the salient features
of this field are: \textit{(i)} it is oriented primarily in the
growth direction; and \textit{(ii)} for each spin subband, the
carrier spin points in the opposite direction on either side of
k-space (about the [00\={1}] axis). The calculated field in
Fig.~\ref{fig:110SpinValveAndOmega}(b) corresponds to the first
heavy hole subband (HH1) in the GaSb/AlSb quantum well within the
proposed spin injector/detector, but these primary characteristics
are reproduced by both valence and conduction states for any (110)
III-V semiconductor quantum well.\cite{notefield}  The crystal
magnetic field in Fig.~\ref{fig:110SpinValveAndOmega}(b)-(c)
differs fundamentally from the clockwise and counter-clockwise
pinwheel structure associated with the Rashba
effect\cite{Ting:2002,Cartoixa:2001,Rashba:1984}; in contrast, for
(110) symmetric heterostructures there is a natural choice of
quantization axis for spin in the growth direction. In this case,
the BIA crystal magnetic field lifts the degeneracy of the
electron spin states but induces only very small precessional
relaxation, resulting in long spin lifetimes. The strong
enhancement of $T_1$ due to the near growth direction orientation
of the BIA crystal magnetic field in (110)-oriented
heterostructures was recently observed in GaAs/AlGaAs quantum
wells\cite{Ohno:1999} and InAs/GaSb
superlattices.\cite{OurCondmat}

Electron spin injection in the proposed device is achieved through
spin-dependent RIT\cite{Ting:2002,Soderstrom:1989}. Resonant
tunneling has been utilized in both
magnetic\cite{Petukhov:2002,Hanbicki:2001} and
nonmagnetic\cite{Voskoboynikov:2000,deAndradaeSilva:1999,Koga:PRL2002,Ting:2002}
spintronics device proposals in recent years.  As shown in the
inset of Fig.~\ref{fig:110SpinFilter}(a), electrons in the
conduction band of the InAs emitter tunnel through the two HH1
spin states of the GaSb quantum well, whose degeneracy is lifted
by BIA, to the InAs 2DEG lateral transport channel. Spin
independent transport in similar RIT structures has been under
investigation for more than a decade because of the advantages of
the broken gap band alignment of the InAs/GaSb heterojunction for
high speed electronic applications.\cite{Soderstrom:1989,Fay:2002}
The calculated HH1 spin splitting in the GaSb quantum well is
shown in Fig.~\ref{fig:110SpinFilter}(a) for a range of well
thicknesses. For thin GaSb layers, the BIA-induced spin splitting
exceeds 30 meV, and is larger than that utilized in similar spin
filter devices based on the Rashba
effect.\cite{Voskoboynikov:2000,deAndradaeSilva:1999,Koga:PRL2002,Ting:2002}
The spin splitting for the first conduction subband of the GaSb
quantum well, which is not used in the present device, is $\leq 20
\%$ smaller for the range of in-plane wavevectors shown and is
considerably larger than the conduction band Rashba spin splitting
in InGaAs/InAlAs
heterostructures.\cite{Grundler:2000,Sato:2001,Nitta:1997,Gui:2000}
These large spin splittings, which reflect the strong spin-orbit
interaction in GaSb, would permit filtering of electron spins at
room temperature with suitably designed emitter and collector
regions.

Since the crystal magnetic field for each of the HH1 spin subbands
reverses sign on either side of the [00\={1}] axis, net spin
injection is achieved by applying a lateral bias along [\={1}10]
using side gates across the InAs emitter, as depicted
schematically in Fig.~\ref{fig:110SpinValveAndOmega}(a). The use
of a lateral bias to generate spin-polarized current was described
previously in analogous resonant tunnel spin filter devices
employing the Rashba effect.\cite{Voskoboynikov:2000,Ting:2002}
Fig.~\ref{fig:110SpinFilter}(b)-(d) shows the calculated valence
subband structure of the GaSb quantum well, together with the
conduction band states of the bulk InAs emitter and the InAs 2DEG
lateral transport channel, which is lightly doped (N$_d$ = 1
$\times$ 10$^{17}$ cm$^{-3}$, fermi energy 30 meV). In
Fig.~\ref{fig:110SpinFilter}(c)-(d), +z (-z) polarized electrons
(+z = [110]) are indicated by the shaded (diagonal striped)
regions.  Electrons that resonantly tunnel through the structure
will emerge in the InAs 2DEG with their spins aligned with the
resonant levels in the quantum
well,\cite{Voskoboynikov:2000,deAndradaeSilva:1999,Ting:2002}
\textit{i.e.}, in the $\pm$z direction. As shown in
Fig.~\ref{fig:110SpinFilter}(c), under the application of a
lateral bias to the bulk InAs emitter, the requirements of
conservation of energy and in-plane momentum for the tunneling
electrons lead to the dominance of tunnel current involving
electrons of one spin (+z is shown: the direction of spin
polarization is determined by the polarity of the lateral bias).
The situation of Fig.~\ref{fig:110SpinFilter}(c) in principle
leads to 100 \% spin-polarized injection, leading to a partially
spin-polarized distribution in the lightly doped InAs 2DEG
channel. For the inter-band spin filter design proposed here, only
a single hole subband is involved in the tunneling process,
avoiding the need to use pauli blocking of collector states to
achieve high spin filtering efficiency.\cite{Ting:2002} The
heavy-hole (HH) light-hole (LH) splitting in the GaSb quantum well
is very large (280 meV), resulting in strongly reduced HH-LH band
mixing and the corresponding suppression of Elliott-Yafet spin
relaxation processes\cite{Elliott:1954,Yafet:1963} for electrons
as they traverse the confined hole states in the GaSb quantum
well. Additionally, a very small bias is required between the
emitter and collector regions, generating negligible associated
Rashba effects.

\begin{figure}[t]\vspace{0pt}
    \includegraphics[width=7.0cm]{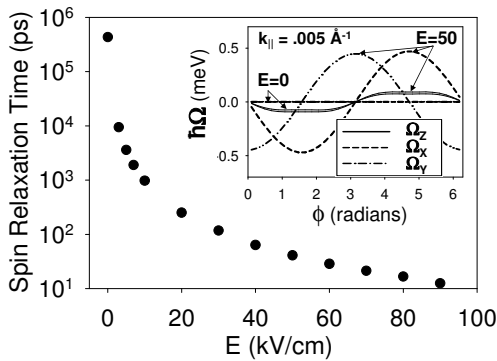}
    \caption{Calculated longitudinal spin relaxation time in the (110) InAs/AlSb quantum well as a function
    of electric field along the growth direction for a lattice temperature of 77K. A momentum relaxation time of 100 fs was assumed. Inset:
     Calculated crystal magnetic field for an applied field of E=0 kV/cm and E=50 kV/cm, illustrating the growth
     of the in-plane Rashba field component with applied bias.}
    \label{fig:BiasDepT1InAs2DEG}
\end{figure}

Spin-dependent resonant tunneling in a second GaSb quantum well
(see Fig.~\ref{fig:110SpinValveAndOmega}(a)) is used for detection
of spin-polarized electrons following lateral transport in the
InAs 2DEG channel. As shown in Fig.~\ref{fig:110SpinFilter}(d), as
long as a net spin polarization persists in the 2DEG at the second
RIT structure, and the range of energies between the fermi levels
of the minority and majority spin electrons is in resonance with
the BIA-split spin levels of the GaSb quantum well (a condition
which may be controlled through appropriate choice of doping level
in the 2DEG channel) a ballistic lateral current will be generated
in the final InAs collector due to the preferential tunneling of
spins on one side of {\bf k}-space relative to the [00\={1}]
direction. In the steady state, this ballistic current will be
detected as a voltage at the side gates across the collector. The
polarity of this voltage indicates the orientation of the initial
spin polarization in the 2DEG. If no spin polarization survives
following transport in the 2DEG, the tunnel current will have
equal contributions on both sides of {\bf k}-space, producing no
voltage across the collector side gates. As this spin detection
scheme relies only on a difference in population between the two
spin states, it does not require ballistic transport in the InAs
2DEG channel.

Spin transistor action is achieved through control over the spin
relaxation time in the 2DEG using the electric field-induced
Rashba
effect.\cite{Rashba:1984,Grundler:2000,Sato:2001,Nitta:1997,Heida:1998,Matsuyama:2000,Lau:JAP2002}
The calculated $T_1$ for electrons in the (110) InAs/AlSb quantum
well is shown in Fig.~\ref{fig:BiasDepT1InAs2DEG} as a function of
the applied electric field.  In the absence of the field, the
precessional spin lifetime is extremely long because both the
injected electron spins and the BIA crystal magnetic field are
oriented primarily in the growth direction.  For comparison, $T_1$
is more than 3 orders of magnitude smaller for the corresponding
(001) InAs/AlSb quantum well,\cite{noteNIA} reflecting the
in-plane orientation of the BIA magnetic field in (001)
heterostructures. $T_1$ in (110) quantum wells is limited only by
a very small, in-plane BIA magnetic field component that
originates from contributions of higher than 3rd order in the
electron wavevector. The applied electric field introduces
structural inversion asymmetry that produces an in-plane Rashba
magnetic field component,\cite{Lau:JAP2002} as shown in the inset
to Fig.~\ref{fig:BiasDepT1InAs2DEG}.  This in-plane magnetic field
induces rapid precessional relaxation of the
growth-direction-polarized electron spins in the InAs 2DEG,
thereby preventing detection at the final RIT spin filter. Because
of the strong spin-orbit interaction in the AlSb barriers and the
small band gap of the type II InAs/AlSb quantum well, the
application of a relatively weak electric field leads to a sharp
decrease in spin lifetime: $T_1$ falls by more than 2 orders of
magnitude for E $<$ 5 kV/cm.  This pronounced sensitivity of $T_1$
to the applied field may permit operation of the spin transistor
proposed here with a low threshold gate voltage, representing a
likely advantage over conventional transistor technology.  It
should be noted that the gate voltage only controls the spin decay
rate associated with precessional relaxation, which is known to
dominate in most III-V
semiconductors\cite{Lau:2001,ALS:book,OpticalOrientationBook}
(including InAs/GaSb/AlSb
heterostructures\cite{Lau:2001,ALS:book,Olesberg:2001,Boggess:2000,Hall:1999})
above 77 K. The residual relaxation rate due to other processes
that may become important with the strong suppression of
precessional relaxation in this (110) 2DEG (such as the
Elliott-Yafet mechanism,\cite{Elliott:1954,Yafet:1963}) will
ultimately determine the requirements for the lateral dimension of
the 2DEG transport channel and the minimum applied field required
to produce the spin transistor action in the device proposed here.

In summary, we have proposed a non-magnetic semiconductor spin
transistor that utilizes the characteristics of BIA effects in
(110) III-V semiconductor quantum wells. Our demonstration that
extremely large spin splittings associated with BIA in 6.1 \AA
\hspace{.03in}semiconductor heterostructures are possible,
together with the long spin lifetimes we calculate for these
structures, illustrates the strong potential for applications of
BIA in a wide range of (110) semiconductor spintronics device
concepts. For example, because of the growth direction orientation
of the electron spins in such structures, vertical emission
spin-polarized light-emitting diodes\cite{Fiederling:1999} may be
realized based on BIA-mediated resonant tunneling spin filters
such as that proposed here.  A non-magnetic semiconductor spin
valve may also be realized using a device similar to that in
Fig.~\ref{fig:110SpinValveAndOmega}(a),\cite{note:spinvalve} in
which the resistivity between the source and drain is controlled
by the relative polarity of a bias applied to the side contacts of
the injection and detection RIT structures.  Due to the large g
factor in InAs, the present device may also find application as a
magnetic field sensor, in which the phase of coherent spin
precession of electrons in the InAs channel in the presence of an
external B field would be indicated by the magnitude and polarity
of the voltage measured at the collector RIT.  We estimate that a
magnetic field as low as a few gauss may be measurable using such
a device.

The authors thank D. Z.-Y. Ting for valuable discussions. This
research is supported by the DARPA MDA972-01-C-0002, DARPA/ARO
DAAD19-01-1-0490 and the Natural Sciences and Engineering Research
Council of Canada.


\begin{references}
\bibitem{notefield} Our calculations indicate that these
characteristics are maintained for electron wavevectors in excess
of 0.1 $\AA^{-1}$.
\bibitem{noteNIA} Our calculation of $T_1$ in the (001)
InAs/AlSb quantum well includes BIA effects only.  Native
interface asymmetry, which does not occur in (110) quantum wells,
will contribute to a further reduction in $T_1$ for the (001)
oriented 2DEG, as discussed in Ref [15].
\end{references}
\end{document}